\documentclass[11pt]{article}
\pdfoutput=1

\usepackage{jheppubmod}
\usepackage{hyperref}
\usepackage{mathtools}
\usepackage{psfrag}
\usepackage{array}
\usepackage{amssymb}
\usepackage{amsmath}
\usepackage{cancel}
\usepackage{braket}
\allowdisplaybreaks
\usepackage{mathrsfs}
\usepackage{wrapfig}
\usepackage{cite}

\usepackage{amsthm}
\usepackage{graphicx}
\usepackage{caption}
\usepackage{paralist}
\usepackage{comment}

\usepackage{epsfig}

\def\be{\begin{equation}}
\def\ee{\end{equation}}
\def\bea{\begin{eqnarray}}

\def\eea{\end{eqnarray}}

\def\half{\frac12}

\def\6{\partial}

 \def\Tr{{\mbox{Tr}}\,}                
        \def\be{\begin{eqnarray}}    \def\ee{\end{eqnarray}}
 \def\bi#1{\begin{itemize}\item[#1]}     \def\ei{\end{itemize}}  
   \def\^#1{\hat{#1}}

   \def\bra{\langle}   

 \def\fract#1#2{{\textstyle{#1\over#2}}}
 \def\ffract#1#2{\raise .2 em\hbox{$\scriptstyle#1$}\kern-.3em/
                 \kern-.2em\lower .15 em \hbox{$\scriptstyle#2$}}
 
 \def\half{\fract12}

\def\bmatrix{\begin{matrix}} \def\ematrix{\end{matrix}} \def\bpmatrix{\begin{pmatrix}}\def\epmatrix{\end{pmatrix}}
\def\bcenter{\begin{center}} \def\ecenter{\end{center}}

\def\lowerheightfig#1#2#3{\(\raise-#1\hbox{\includegraphics[height=#2]{#3}}\)}
\def\lowerwidthfig#1#2#3{\(\raise-#1\hbox{\includegraphics[width=#2]{#3}}\)}

\title{The Black Hole S-Matrix from Quantum Mechanics}
\author{Panagiotis Betzios, }
\author{Nava Gaddam, }
\author{Olga Papadoulaki} 
\affiliation{Institute for Theoretical Physics and Center for Extreme Matter and Emergent Phenomena,
Utrecht University, 3508 TD Utrecht, The Netherlands.}
\emailAdd{P.Betzios@uu.nl}
\emailAdd{gaddam@uu.nl}
\emailAdd{O.Papadoulaki@uu.nl}

\date{}

\abstract{We revisit the old black hole S-Matrix construction and its new partial wave expansion of 't Hooft. Inspired by old ideas from non-critical string theory \& $c=1$ Matrix Quantum Mechanics, we reformulate the scattering in terms of a quantum mechanical model\textemdash of waves scattering off inverted harmonic oscillator potentials\textemdash that exactly reproduces the unitary black hole S-Matrix for all spherical harmonics; each partial wave corresponds to an inverted harmonic oscillator with ground state energy that is shifted relative to the s-wave oscillator. Identifying a connection to 2d string theory allows us to show that there is an exponential degeneracy in how a given total initial energy may be distributed among many partial waves of the 4d black hole.}
\begin{document}
\maketitle

\section{Introduction}
The work of Bekenstein \cite{Bekenstein:1973ur} and Hawking \cite{Hawking:1974sw} has spurred extensive research on the so-called `information paradox'. The premise of the paradox is that in a collapse of matter forming a black hole, the intermediate state post-collapse is a black hole that can be characterized by a small number of physical parameters (mass, charge, angular momentum, etc.). A semi-classical calculation as the one Hawking originally did, however, suggests that black holes radiate as black bodies, namely with a thermal spectrum. This seems to suggest a gross violation of unitary evolution as all information about the exact in-state that went into forming the black hole appears to have been lost after its evaporation. \\

The emergence of string theory, holography \cite{'tHooft:1996tq,'tHooft:1984re,Susskind:1994vu} and gauge-gravity duality \cite{Maldacena:1997re,Gubser:1998bc,Witten:1998qj} has shed significant light on this problem. In fact, it is often claimed that if one were to believe gauge-gravity duality, the paradox is solved `in-principle' as the boundary theory is unitary by construction and the duality states an equivalence (at the level of partition functions) between the gravitational and boundary field theories. Nevertheless, the strength of this claim is sometimes questioned \cite{Mathur,Mathur:2009hf}, and even within the best understood examples of gauge-gravity duality, there is no general consensus on the exact process of information retrieval. Furthermore, the best understood examples of the said duality, while providing for a very useful toolbox, typically involve bulk space-times with a negative cosmological constant and are far from the real world. Technology at this stage is far from established to reliably understand more realistic space-times. Additionally, why intricate details of string theory or the duality may be absolutely necessary for our understanding of the evolution of general gravitational dynamics is not apparent. While the fuzzball program \cite{Mathur:2005zp,Bena:2007kg,Balasubramanian:2008da,Skenderis:2008qn,Mathur:2008nj} provides some arguments for why stringy details may be important, it is fair to say that there is no general consensus on the matter.  \\ 

Years before gauge-gravity duality was proposed and was seen as a possible resolution to the information paradox, there was an alternative suggestion by 't Hooft \cite{'tHooft:1996tq,'tHooft:1991bd,'tHooft:1992zk}. The proposal was to consider particles of definite momenta `scattering' off a black-hole horizon. These particles were to impact the out-going Hawking quanta owing to their back-reaction on the geometry. With the knowledge that the black hole is made out of a large, yet \textit{finite}, number of in-states, one may scatter particles of varying momenta repeatedly, until all in-states that may have made up the black hole have been exhausted. This led to a construction of an S-Matrix that maps in to out states. This matrix was shown to be unitary. A further advancement for spherically symmetric horizons was made recently \cite{Hooft:2015jea,Hooft:2016itl,Hooft:2016cpw}, where a partial wave expansion allowed for an explicit writing of the S-Matrix for each spherical harmonic. However, this construction has its own short-comings. It presumes that the S-Matrix can be split as
\begin{equation}\label{eqn:SMatrixsplit}
S_{\text{total}} ~ = ~ S_{-\infty} ~ S_{\text{horizon}} ~ S_{+\infty} \, ,
\end{equation}
where $S_{\pm\infty}$ correspond to matrices that map asymptotic in-states to in-going states near the horizon and outgoing states near the horizon to asymptotic out-states respectively. And $S_{\text{horizon}}$ is the S-Matrix that captures all the dynamics of the horizon. Whether such an arbitrarily near-horizon region captures all the dynamics of the black hole is not entirely clear. The construction is also done in a ``probe-limit" in that the back-reaction is not taken to impact the mass of the black hole. Only its effect on outgoing particles is captured. Furthermore, throwing a particle into a black hole is not an exactly spherically symmetric process. While a non-equilibrium process initiated by the in-going particle does break spherical symmetry, it may be expected that the black hole settles down into a slightly larger, spherically symmetric solution after some characteristic time-scale that depends on the interactions between the various degrees of freedom that make up the black hole. This scattering can be decomposed into partial waves. And the different waves are shown to evolve independently. However, one might expect that the partial waves are not independent and that they could indeed ``interact" in a generic evolutionary process; it is not clear how one may incorporate this interaction in this construction. This is related to the known limitation of the back-reaction calculations ignoring transverse effects \cite{Aichelburg:1970dh,Dray:1984ha}, which grow in increasing importance as we approach Planckian scales. Finally, while it may not be a fundamental difficulty, the splitting of the wave function via \eqref{eqn:SMatrixsplit} needs further investigation. \\

As is evident from the above, it is surprisingly easy to criticize even the most promising approaches to quantum black hole physics. In this article, we seek to address some of the criticisms of the S-Matrix approach to quantum black holes. Inspired by old ideas from non-critical 2d string theory \cite{Klebanov:1991qa,Moore1992b,Schoutens:1993hu,Verlinde:1993sg,Alexandrov:2002fh,Karczmarek:2004bw,Friess:2004tq,Maldacena:2005he}, we construct a theory describing a \textemdash  collection of partial waves scattering in an inverted harmonic oscillator potential\textemdash that exactly reproduces the S-Matrix of 't Hooft; the inverted potential arises naturally to allow for scattering states, as opposed to bound states in a conventional harmonic oscillator. The intrinsically quantum nature of the model dispenses with the critique that the S-Matrix of 't Hooft is a `classical' one. With any toy model, it may be hard to establish the validity of its applicability to black hole physics. However, in our construction, all the observables (S-Matrix elements) are exactly identical to those of 't Hooft's S-Matrix; thereby avoiding any ambiguity of its validity. Furthermore, we observe that in-states must contain an approximately constant number density over a wide range of frequencies in order for the scattered out-states to appear (approximately) thermal; this condition was also noted in the 2d string theory literature. Finally, and perhaps most significantly, we show that our model captures an exponentially growing degeneracy of states. \\

It may be added that aside from the approaches mentioned earlier, there have been many attempts to construct toy-models to study black hole physics \cite{Callan:1992rs,Gibbons:1998fa,Kazakov2002,Magan:2016ojb,Jansen:2016zai,Banerjee:2016mhh}. The hope being that `good' toy models teach us certain universal features of the dynamics of black hole horizons. \\

This article is organized as follows. In the section \ref{sec:macroSMatrix}, we briefly review gravitational back-reaction and 't Hooft's S-Matrix construction along with its partial wave expansion. Our derivation is slightly different to the one of 't Hooft \cite{Hooft:2015jea} in that our derivation relies only on the algebra associated to the scattering problem. Therefore, the `boundary conditions' of the effective bounce, as was imposed in 't Hooft's construction is built in from the start via the back-reaction algebra \eqref{eqn:macroalgebra}. In section \ref{sec:micromodel}, we present our model and compute the corresponding scattering matrix to show that it explicitly matches the one of 't Hooft. In section \ref{sec:collapse}, we make an estimate of the high energy behaviour of the total density of states to argue that the model indeed describes the existence of an intermediate black-hole state. We conclude with a discussion and some future perspectives in \ref{sec:discussion}. \\

\paragraph{A brief summary of results: } There are two main results of this work: one is a re-writing the degrees of freedom associated to 't Hooft's black hole S-Matrix in terms of partial waves scattering on an inverted harmonic oscillator potential; this allows us to write down the corresponding Hamiltonian of evolution explicitly. The second, related result is an identification of a connection to 2d string theory which in turn allows us to show that there is an exponential degeneracy of how a given total initial energy may be distributed among many partial waves of the 4d black hole; much as is expected from the growth of states associated to black hole entropy. At various points in Sections \ref{sec:micromodel} and \ref{sec:collapse}, we review some aspects of matrix models and 2d string theory in detail. While we expect some consequences for these theories based on our current work, we do not have any new results within the framework of 2d black holes or matrix models in this paper.

\section{Back-reaction and the Black Hole S-Matrix}\label{sec:macroSMatrix}

Consider a vacuum solution to Einstein's equations of the form:
\begin{equation}\label{eqn:genericmetric}
ds^2 ~ = ~ 2 A\left(u^+,u^-\right) du^+ \, du^- + g\left(u^+,u^-\right) \, h\left(\Omega\right) d\Omega^2 \, ,
\end{equation}
where $u^+, u^-$ are light-cone coordinates, $A\left(u^+,u^-\right)$ and $g\left(u^+,u^-\right)$ are generic smooth functions of those coordinates and $h\left(\Omega\right)$ is the metric tensor depending on only the $(d-2)$ transverse coordinates $\Omega$. It was shown in \cite{Dray:1984ha} that an in-going massless particle with momentum $p^-$ induces a shock-wave at its position specified by $\Omega$ and $u^+$ (it impacts the future horizon when it crosses $u^-=0$). The shock-wave was shown to change geodesics such that out-going massless particles feel a ``kick" \textemdash of the form $u^- \rightarrow u^- + 8 \pi G \, p^-_{\text{in}} \hat{f}\left(\Omega,\Omega^\prime\right)$ \textemdash in their trajectories, that start initially at $u^-=0$. The ``kick" is parametrised by $\hat{f}$ that depends on the spacetime in question. If we were to associate a putative S-Matrix to the dynamics of the black hole, the said back-reaction may be attributed to this S-Matrix in the following manner. Consider a generic in-state $| \text{in}_0 \rangle$ that collapsed into a black hole and call the corresponding out-state after the complete evaporation of the black hole $|\text{out}_0 \rangle$. The S-Matrix maps one into the other via: $S \, |\text{in}_1 \rangle = | \text{out}_1 \rangle$. Now the back-reaction effect may be treated as a tiny modification of the in-state as $| \text{in}_0 \rangle \rightarrow |\text{in}_0  + \delta p^-_{\text{in}}\left(\Omega\right) \rangle $, where $\delta p^-_{\text{in}}\left(\Omega\right)$ is the momentum of an in-going particle at position $\Omega$ on the horizon. Consequently, the action of the S-Matrix on the modified in-state results in a different out-state which is acted upon by an operator that yields the back-reacted displacement:
\begin{equation}
S \, | \text{in}_0 + \delta p^-_{\text{in}}\left(\Omega\right) \rangle ~ = ~ e^{- i \delta p^+_{\text{out}}\left(\Omega^\prime\right) \delta u^-_{\text{out}}} |\text{out}_0 \rangle \, ,
\end{equation}
where the operator acting on the out-state above is the `displacement' operator written in Fourier modes. Now, we may repeat this modification arbitrarily many times. This results in a cumulative effect arising from all the radially in-going particles with a distribution of momenta on the horizon. Therefore, writing the new in- and out-states\textemdash with all the modifications included\textemdash as $|\text{in} \rangle$ and $| \text{out}\rangle$ respectively, we have
\begin{equation}
\Bra{\text{out}}S \Ket{\text{in}} ~ = ~ \Bra{\text{out}_0}S\Ket{\text{in}_0} \, \exp \left[- i 8 \pi G \int d^{d-2}\Omega^\prime \, p^+_{\text{out}}\left(\Omega^\prime\right) \, \hat{f}\left(\Omega,\Omega^\prime\right) \, p^-_{\text{in}}\left(\Omega\right)\right] \, .
\end{equation}
Should we now \textit{assume} that the Hilbert space of states associated to the black-hole is completely spanned by the in-going momenta and that the Hawking radiation is entirely spanned by the out-state momenta, we are naturally led to a unitary S-Matrix given by
\begin{equation}
\Bra{p^+_\text{out}}S\Ket{p^-_\text{in}} ~ = ~ \exp \left[- i 8 \pi G \int d^{d-2}\Omega^\prime \, p^+_{\text{out}}\left(\Omega^\prime\right) \, \hat{f}\left(\Omega,\Omega^\prime\right) \, p^-_{\text{in}}\left(\Omega\right)\right] \, .
\end{equation}
There is an overall normalization factor (vacuum to vacuum amplitude) that is undetermined in this construction. The assumption that the black hole Hilbert space of states is spanned entirely by the in-state momenta $p^-_{\text{in}}$ is equivalent to postulating that the said collection of radially in-going, gravitationally back-reacting particles could collapse to form a black hole. While this may seem a reasonable assumption, it is worth emphasizing that there is no evidence for this at the level of the discussion so far. We have not modeled a collapsing problem. We will see in Section \ref{sec:collapse} that our proposed model in Section \ref{sec:micromodel} provides for a natural way to study this further. And significantly, we give non-trivial evidence that the derived S-Matrix could possibly model a collapsing black-hole. 

\subsection{Derivation of the S-Matrix}
We now return to the back-reaction effect at a semi-classical level in order to derive an explicit S-Matrix using a partial wave expansion in a spherically symmetric problem. For the back-reacted metric\textemdash after incorporating the shift $u^- \rightarrow u^- + f\left(\Omega,\Omega^\prime\right)$ into \eqref{eqn:genericmetric}\textemdash to still satisfy Einstein's equations of motion, the following conditions need to hold at $u^-=0$ \cite{Dray:1984ha}:
\begin{align}\label{eqn:backreactionconds.}
\dfrac{A\left(u^{+,-}\right)}{g\left(u^{+,-}\right)} \bigtriangleup_\Omega f\left(\Omega,\Omega^\prime\right) - \left(\dfrac{d-2}{2}\right)\dfrac{\partial_{u^+} \partial_{u^-} g\left(u^{+,-}\right)}{g\left(u^{+,-}\right)} f\left(\Omega,\Omega^\prime\right) ~ &= ~ 8 \, \pi \, p^-_{\text{in}} \,  A\left(u^{+,-}\right)^2 \, \delta^{(d-2)}\left(\Omega,\Omega^\prime\right) \nonumber \\
\partial_{u^-} A\left(u^{+,-}\right) ~ &= ~ 0 ~ = ~ \partial_{u^-} g\left(u^{+,-}\right)  \, ,
\end{align}
where $\bigtriangleup_\Omega$ is the Laplacian on the $(d-2)$-dimensional metric $h\left(\Omega\right)$. We concern ourselves with the Schwarzschild black-hole, written in Kruskal-Szekeres coordinates as
\begin{equation}\label{eqn:metricinKruskal}
ds^2 ~ = ~ - \dfrac{32 \, G^3 \, m^3}{r} e^{-r/2Gm} du^+ \, du^- + r^2 d\Omega^2 \, .
\end{equation}
For the above metric \eqref{eqn:metricinKruskal}, at the horizon $r = R = 2 G m$, the conditions \eqref{eqn:backreactionconds.} were shown \cite{Dray:1984ha} to reduce to
\begin{equation}\label{eqn:laplacianonshift}
\bigtriangleup_S \left(\Omega\right) f\left(\Omega,\Omega^\prime\right) ~ \coloneqq ~ \left(\bigtriangleup_\Omega - 1\right) f\left(\Omega,\Omega^\prime\right) ~ = ~ - \kappa \, \delta^{(d-2)}\left(\Omega,\Omega^\prime\right) \, ,
\end{equation}
with the implicit dependence of $r$ on $u^+$ and $u^-$ given by
\begin{equation}
u^+ \, u^- ~ = ~ \left(1 - \dfrac{r}{2 G m}\right) e^{-r/2 G m} \, ,
\end{equation}
and $\kappa = 2^4 \, \pi \, e^{-1} \, G \, R^2 \, p^-_{\text{in}}$. These seemingly ugly coefficients may easily be absorbed into the stress-tensor on the right hand side of the Einstein's equations. Now, the cumulative shift experienced by an out-going particle, say $u^-_{\text{out}}$, is given by a distribution of in-going momenta on the horizon
\begin{align}\label{eqn:cumulativeshift1}
u^-_{\text{out}}\left(\Omega\right) ~ &= ~ 8 \pi G R^2 \, \int d^{d-2} \Omega^\prime \, \tilde{f}\left(\Omega,\Omega^\prime\right) \, p^-_{\text{in}}\left(\Omega^\prime\right) \, ,
\end{align}
where $\kappa \, \tilde{f}\left(\Omega,\Omega^\prime\right) = f\left(\Omega,\Omega^\prime\right)$. Similarly, we have the complementary relation for the momentum of the out-going particle, say $p^+_{\text{out}}$ given in terms of the position $u^+_{\text{in}}$ of the in-going particle:
\begin{align}\label{eqn:cumulativeshift2}
u^+_{\text{in}}\left(\Omega\right) ~ &= ~ -8 \pi G R^2 \, \int d^{d-2} \Omega^\prime \, \tilde{f}\left(\Omega,\Omega^\prime\right) \, p^+_{\text{out}}\left(\Omega^\prime\right) 
\end{align}
The expressions \eqref{eqn:cumulativeshift1} and \eqref{eqn:cumulativeshift2} may be seen as `boundary conditions' of an effective bounce off the horizon. However, this intuition is rather misleading and we will refrain from this line of thought. Nevertheless, what is striking to note is that the momentum of the in-state is encoded in the out-going position of the Hawking radiation while the position of the in-state is encoded in the momentum of the out-going Hawking state! However, so far, the quantities $u^\pm_{\text{in/out}}$ are dimensionless while $p^\mp_{\text{in/out}}$ are densities of momenta with mass dimensions four. Therefore, to appropriately interpret these as positions and momenta, we rescale them as $u^\pm_{\text{in/out}} \rightarrow R u^\pm_{\text{in/out}}$ and $p^\mp_{\text{in/out}} \rightarrow R^{-3} p^\mp_{\text{in/out}}$ \cite{Hooft:2016itl}. Notwithstanding this rescaling, we continue to use the same labels for the said quantities in order to avoid clutter of notation. Now, using the canonical commutation relations, respectively, for the out and in particles\footnote{To avoid clutter in notation, we drop the in/out labels on positions and momenta of particles. $u^+$ and $u^-$ always refer to ingoing/outgoing positions, respectively. Consequently, $p^-$ and $p^+$ are always associated with ingoing/outgoing momenta, respectively.} 
\begin{equation}
\left[\hat{u}^-\left(\Omega\right),\hat{p}^+\left(\Omega^\prime\right)\right] ~ = ~ \left[\hat{u}^+\left(\Omega\right),\hat{p}^-\left(\Omega^\prime\right)\right] ~ = ~ i \, \delta^{(d-2)}\left(\Omega-\Omega^\prime\right) \, ,
\end{equation}
we may derive the algebra associated to the black hole scattering. We do this in a partial wave expansion\textemdash in four dimensions\textemdash as
\begin{equation}
\hat{u}^\pm \left(\Omega\right) ~ = ~ \sum_{lm} \hat{u}^\pm_{lm} \, Y_{lm} \left(\Omega\right)  \quad \text{and} \quad \hat{p}^\pm \left(\Omega\right) ~ = ~ \sum_{lm} \hat{p}^\pm_{lm} \, Y_{lm} \left(\Omega\right) \, .
\end{equation}
Working with these eigenfunctions of the two-sphere Laplacian and using \ref{eqn:laplacianonshift} we can write the back-reaction equations \eqref{eqn:cumulativeshift1} and \eqref{eqn:cumulativeshift2} as
\begin{equation}\label{eqn:partialwavebackreaction}
\hat{u}^\pm_{lm} ~ = ~ \mp \dfrac{8 \pi G}{R^2\left(l^2 + l + 1\right)} \hat{p}^\pm_{lm} ~ \eqqcolon ~ \mp \lambda \, \hat{p}^\pm_{lm} \, .
\end{equation}
In terms of these partial waves, we may now write the scattering algebra as
\begin{align}\label{eqn:macroalgebra}
\left[\hat{u}^\pm_{lm},\hat{p}^\mp_{l^\prime m^\prime}\right] ~ &= ~ i \delta_{l l^\prime} \delta_{m m^\prime} \\
\left[\hat{u}^+_{lm},\hat{u}^-_{l^\prime m^\prime}\right] ~ &= ~ i \, \lambda \, \delta_{l l^\prime} \delta_{m m^\prime}   \\
\left[\hat{p}^+_{lm},\hat{p}^-_{l^\prime m^\prime}\right] ~ &= ~ - \dfrac{i}{\lambda} \, \delta_{l l^\prime} \delta_{m m^\prime}
\end{align}
A few comments are now in order. Since the different spherical harmonics do not couple in the algebra, we will drop the subscripts of $l$ and $m$ from here on. Furthermore, we see that the shift-parameter $\lambda$ `morally' plays the role of Planck's constant $\hbar$, but one that is now $l$ dependent. Moreover, we see that wave-functions described in terms of four phase-space variables are now pair-wise related owing to the back-reaction \eqref{eqn:partialwavebackreaction}. Finally, it is important to note that each partial wave does not describe a single particle but a specific profile of a density of particles. For instance, the $s$-wave with $l=0$ describes a spherically symmetric density of particles. \\

Since the operators $\hat{u}^\pm$ and $\hat{p}^\pm$ obey commutation relations associated to position and momentum operators, we see that the algebra may be realized with $\hat{u}^- = - i \lambda \partial_{u^+}$ in the $u^+$ basis and $\hat{u}^+ = i \lambda \partial_{u^-}$ in the $u^-$ basis. A similar realization is evident for the momentum operators. Moreover, we may now define the following inner-products on the associated Hilbert space of states that respect the above algebra:
\begin{align}
\Braket{u^\pm|p^\mp} ~ &= ~ \dfrac{1}{\sqrt{2 \pi}} \exp\left(i u^\pm p^\mp\right) \\ 
\Braket{u^+|u^-} ~ &= ~ \dfrac{1}{\sqrt{2 \pi \lambda}} \exp\left(i \frac{u^+ u^-}{\lambda}\right) \label{eqn:upmscattering}\\
\Braket{p^+|p^-} ~ &= ~ \sqrt{\dfrac{\lambda}{2 \pi}} \exp\left(i \lambda p^+ p^-\right)
\end{align}
Using \eqref{eqn:upmscattering}, for instance, we may write the out-going wave-function\textemdash travelling along the coordinate $u^-$ after scattering\textemdash in terms of the in-going one travelling along $u^+$ as
\begin{equation}\label{eqn:scatwavefn}
\Braket{u^-|\psi} ~ \eqqcolon ~ \psi^{\text{out}}\left(u^-\right) ~ = ~ \int_{-\infty}^{\infty} \, \dfrac{d u^+}{\sqrt{2 \pi \lambda}} \, \exp\left(-i \frac{u^+ u^-}{\lambda}\right) \, \psi^{\text{in}}\left(u^+\right) \, . 
\end{equation}
One can immediately see that this mapping is Unitary just being a fourier transform.
To derive another useful form of the S-Matrix associated to the scattering, we first move to Eddington-Finkelstein coordinates:
\begin{equation}
u^+ ~ = ~ \alpha^+ \, e^{\rho^+} \, ,\quad u^- ~ = ~ \alpha^- \, e^{\rho^-} \, , \quad \, p^+ ~ = ~ \beta^+ \, e^{\omega^+} \quad \text{and} \quad p^- ~ = ~ \beta^- \, e^{\omega^-}
\end{equation}
where $\alpha^\pm = \pm 1$ and $\beta^\pm = \pm 1$ to account for both positive and negative values of the phase space coordinates $u^+$, $u^-$, $p^+$ and $p^-$. The normalization of the wave-function as
\begin{align}
1 ~ &= ~ \int_{-\infty}^\infty \left|\psi\left(u^+\right)\right|^2 \, d u^+ \nonumber \\
&= ~ \int_{-\infty}^0 \left|\psi\left(u^+\right)\right|^2 \, du^+ ~ + ~ \int_{0}^{\infty} \left|\psi\left(u^+\right)\right|^2 \, d u^+ \nonumber \\
&= ~ - \int_{\infty}^{-\infty} \left|\psi^+\left(- e^{\rho^+}\right)\right|^2 e^{\rho^+} \, d\rho^+ ~ + ~ \int_{-\infty}^{\infty} \left|\psi^+\left(+ e^{\rho^+}\right)\right|^2 e^{\rho^+} \, d\rho^+ \nonumber \\
&= ~ \sum_{\alpha = \pm} \, \int_{-\infty}^{\infty} \left|\psi^+\left(\alpha e^{\rho^+}\right)\right|^2 e^{\rho^+} \, d\rho^+
\end{align}
suggests the following redefinitions for the wave-function in position and momentum spaces
\begin{align}
\psi^\pm\left(\alpha^\pm e^{\rho^\pm}\right) ~ &= ~ e^{-\rho^\pm/2} \, \phi^\pm\left(\alpha^\pm,\rho^\pm\right) \quad \& \quad \tilde{\psi}^\pm\left(\beta^\pm e^{\omega^\pm}\right) ~ = ~ e^{-\omega^\pm/2} \, \tilde{\phi}^\pm\left(\beta^\pm,\omega^\pm\right) \, .
\end{align}
Therefore, using \eqref{eqn:scatwavefn}, we may write $\phi^{\text{out}}\left(\alpha^-, \rho^-\right)$ as:
\begin{align}\label{eqn:fourier1}
\phi^\text{out}\left(\alpha^-, \rho^-\right) ~ &= ~ \dfrac{1}{\sqrt{2 \pi \lambda}} \int_{-\infty}^{\infty} du^+ \, e^{\frac{\rho^+ + \rho^-}{2}} \, \exp\left(-i \frac{u^+ u^-}{\lambda}\right) \, \phi^{\text{in}}\left(\alpha^+,\rho^+\right) \nonumber \\
&= ~ \sum_{\alpha^+ = \pm} \int_{-\infty}^{\infty} \dfrac{du^+ }{\sqrt{2 \pi}} e^{\frac{\rho^+ + \rho^- - \log\lambda}{{2}}} \exp\left(- i \alpha^+ \alpha^- e^{\rho^+ + \rho^- - \log\lambda}\right) \, \phi^{\text{in}}\left(\alpha^+,\rho^+\right) \nonumber \\
&= \sum_{\alpha^+ = \pm} \int_{-\infty}^{\infty} \dfrac{dx}{\sqrt{2 \pi}} \, \exp\left(\frac{x}{2} - i \alpha^+ \alpha^- e^x\right) \, \phi^{\text{in}}\left(\alpha^+,x + \log\lambda - \rho^-\right) \, ,
\end{align}
where in the last line, we introduced $x\coloneqq \rho^+ + \rho^- - \log\lambda$. This equation may be written in matrix form as
\begin{equation}\label{eqn:SMatrix1}
\left( \begin{array}{c}
\phi^{\text{out}}\left(+,\rho^-\right) \\
\phi^{\text{out}}\left(-,\rho^-\right) \end{array} \right) ~ = ~ \int_{-\infty}^\infty \, dx \left( \begin{array}{cc}
A\left(+,+,x\right) & A\left(+,-,x\right) \\
A\left(-,+,x\right) & A\left(-,-,x\right) \end{array} \right) \left( \begin{array}{c}
\phi^{\text{in}}\left(+,x + \log\lambda - \rho^-\right) \\
\phi^{\text{in}}\left(-,x + \log\lambda - \rho^-\right) \end{array} \right)
\end{equation}
where we have defined the quantity
\begin{equation}\label{eqn:Ax}
A\left(\gamma,\delta,x\right) ~ \coloneqq ~ \dfrac{1}{\sqrt{2\pi}} \, \exp\left(\frac{x}{2} - i \, \gamma \, \delta \, e^x\right) \, ,
\end{equation}
with $\gamma = \pm$ and $\delta = \pm$. This integral equation may further be simplified by moving to Rindler plane waves:
\begin{align}
\phi^{\text{out}}\left(\pm,\rho^-\right) ~ &= ~ \dfrac{1}{\sqrt{2 \pi}} \int_{-\infty}^{\infty} dk_- \, \phi^{\text{out}}\left(\pm,k_-\right) \, e^{i k_- \rho^-} \\
\phi^{\text{in}}\left(\pm,x + \log\lambda - \rho^-\right) ~ &= ~ \dfrac{1}{\sqrt{2 \pi}} \int_{-\infty}^{\infty} \, d k_{\tilde{x}} \, \phi^{\text{in}}\left(\pm,k_{\tilde{x}}\right) e^{-ik_{\tilde{x}} \left(x + \log\lambda - \rho^-\right)} \\
A\left(\gamma,\delta,x\right) ~ &= ~ \dfrac{1}{\sqrt{2 \pi}} \int_{-\infty}^{\infty} \, dk_x \, A\left(\gamma,\delta,k_x\right) e^{i k_x \, x}
\end{align}
This allows us to write the above matrix equation \eqref{eqn:SMatrix1} as
\begin{equation}
\left( \begin{array}{c}
\phi^{\text{out}}\left(+,k\right) \\
\phi^{\text{out}}\left(-,k\right) \end{array} \right) ~ = ~ e^{-ik \log\lambda} \left( \begin{array}{cc}
A\left(+,+,k\right) & A\left(+,-,k\right) \\
A\left(-,+,k\right) & A\left(-,-,k\right) \end{array} \right) \left( \begin{array}{c}
\phi^{\text{in}}\left(+,k\right) \\
\phi^{\text{in}}\left(-,k\right) \end{array} \right)
\end{equation}
where $A\left(\gamma,\delta,k\right)$ can be computed from the inverse Fourier transform of \eqref{eqn:Ax} using a coordinate change $y = e^x$ and the identity
\begin{equation}
\int_0^{\infty} \, dy \, e^{i\sigma y} y^{-ik-\frac{1}{2}} ~ = ~ \Gamma\left(\dfrac{1}{2} - ik\right) \, e^{i\sigma\frac{\pi}{4}} \, e^{k\sigma\frac{\pi}{2}} \, , \quad \text{where} \quad \sigma = \pm \, .
\end{equation}
Carrying out this computation, we find the following S-Matrix: 
\begin{align}\label{eqn:SMatrix2}
S\left(k_l,\lambda_l\right) ~ &= ~ e^{-ik_l \log\lambda_l} \left( \begin{array}{cc}
A\left(+,+,k_l\right) & A\left(+,-,k_l\right) \\
A\left(-,+,k_l\right) & A\left(-,-,k_l\right) \end{array} \right) \nonumber \\
&= ~ \dfrac{1}{\sqrt{2\pi}} \Gamma\left(\dfrac{1}{2} - ik_l\right) e^{-ik_l \log\lambda_l} \left( \begin{array}{cc}
e^{-i\frac{\pi}{4}} \, e^{-k_l\frac{\pi}{2}} & e^{i\frac{\pi}{4}} \, e^{k_l\frac{\pi}{2}} \\
e^{i\frac{\pi}{4}} \, e^{k_l\frac{\pi}{2}} & e^{-i\frac{\pi}{4}} \, e^{-k_l\frac{\pi}{2}} \end{array} \right)
\end{align}
In this expression, we have reinstated a subscript on $k$ and $\lambda$ to signify that they depend on the specific partial wave in question. One may additionally diagonalize this matrix by noting that 
\begin{equation}
A\left(+,+,k\right) ~ = ~ A\left(-,-,k\right) \quad \text{and} \quad A\left(+,-,k\right) ~ = ~ A\left(-,+,k\right) \, .
\end{equation}
With this observation, we see that the diagonalization of the S-Matrix is achieved via the redefinitions
\begin{align}
\phi^+_1\left(k\right) ~ &= ~ \phi^+\left(+,k\right) + \phi^+\left(-,\rho^+\right) \, , ~~ \qquad \phi^+_2\left(k\right) ~ &&= ~ \phi^+\left(+,k\right) - \phi^+\left(-,k\right) \nonumber \\
A_1\left(k\right) ~ &= ~ A\left(+,+,k\right) + A\left(+,-,k\right) \,, \qquad A_2\left(k\right) ~ &&= ~ A\left(+,+,k\right) - A\left(+,-,k\right) \, .
\end{align}
It may be additionally checked that this matrix is unitary. As already mentioned, while it may not be clear whether this matrix is applicable to the formation and evaporation of a physical black hole, a conservative statement that can be made with certainty is the following: all information that is thrown into a large black hole is certainly recovered in its entirety, at least when the degrees of freedom in question are positions and momenta. It would be interesting to generalize this to degrees of freedom carrying additional conserved quantities like electric charge, etc. On the other hand, there is a certain property of the S-Matrix that may be puzzling at first sight. Positive Rindler energies $k$ imply that the off-diagonal elements in the S-Matrix are dominant with exponentially suppressed diagonal elements. While negative Rindler energies reverse roles. One way to interpret this feature is to think of an eternal black hole where dominant off-diagonal elements suggest that information about in-going matter from the right exterior is carried mostly by out-going matter from the left exterior. However, in a physical collapse, there is only one exterior. It has been suggested by 't Hooft that one must make an antipodal mapping between the two exteriors to make contact with the one-sided physical black hole; we discuss this issue in Section \ref{sec:discussion}.

\section{The model}\label{sec:micromodel}

Asking two simple questions allows us to almost entirely determine a quantum mechanical model that corresponds to the black hole scattering matrix of the previous section. The first question is `what kind of a quantum mechanical potential allows for scattering states?' The answer is quite simply that it must be an unstable potential. The second question is `what quantum mechanical model allows for energy eigenstates that resemble those of Rindler space?' The answer, as we will show in this section, is a model of waves scattering off an inverted harmonic oscillator potential. Using this intuition, we will now construct the model and show that it explicitly reproduces the desired S-Matrix. Having constructed the model, we will then proceed to compare it to 2d string theory models. The construction of our model and intuition gained from a comparison to 2d string theory/matrix quantum mechanics models \cite{Moore1992b,Alexandrov:2002fh,Maldacena:2005he} allows us to study time delays and degeneracy of states in the next section. \\

Inverted quadratic potentials, at a classical level, fill up phase space with hyperbolas as opposed to ellipses as in the case of standard harmonic oscillator potentials. Since we have a tower of 4d partial waves in the black hole picture, each of them results in a phase space of position and momentum and consequently a collection of inverted harmonic oscillators, one for each partial wave. Since the black hole scattering of 't Hooft mixes positions and momenta, we are naturally led to consider the description of scattering in phase space. 

\subsection{Construction of the model}

We first start with a phase space parametrized by variables $x_{l m}$ and $p_{l m}$. To implement the appropriate scattering off the horizon, we start with the same black hole scattering algebra: $[\hat x_{l m} , \hat p_{l' m'} ]= i  \lambda \delta_{m m'} \delta_{l l'}$, with $\lambda = c/\left(l^2+l+1\right)$ with $c=8 \pi G/R^2$. We will return to how this parameter might naturally arise in a microscopic setting in Section \ref{sec:discussion}. Standard bases of orthonormal states are $|x; l, m \rangle$ and $|p; l, m \rangle$; these are coordinate and momentum eigenstates respectively, with 
\begin{equation}
\langle l, m ; x |p ; l, m \rangle ~ = ~ \dfrac{1}{\sqrt{2 \pi \lambda}} \,  e^{ipx/ \lambda} \, \delta_{m m'} \delta_{l l'} \, . 
\end{equation}
Since our interest is in the scattering of massless particles, it will turn out to be convenient to use light-cone bases $|u^\pm ; l, m \rangle$ which are orthonormal eigenstates of the light-cone operators:
\begin{equation}
\hat u^\pm_{l m} ~ = ~ \dfrac{\hat{p}_{l m} \pm \hat{x}_{l m}}{\sqrt 2} \qquad \text{and} \qquad \left[\hat u^+_{l m} ,\hat u^-_{l' m'} \right] ~ = ~ i \lambda \delta_{l l'}\delta_{m m'} \, .
\end{equation}
While they look similar to creation and annihilation operators of the ordinary harmonic oscillator, $\hat u^\pm$  are in truth hermitian operators themselves; and are not hermitian conjugate to each other. Therefore, the states $|u^\pm ; {l, m}\rangle$ are reminiscent of coherent states. These plus and minus bases will be useful in describing the in and outgoing states of the upside down harmonic
oscillator. For definiteness, we will choose for the ingoing states to be described in terms of the $u^+_{l, m}$ basis while for the outgoing ones to be in terms of the $u^-_{l, m}$ basis. As in the previous section, we will work in the simplification where different oscillators (partial waves) do not interact and will therefore omit the partial wave labels in all places where they do not teach us anything new. Furthermore, as before, from the commutation relations we may define the following inner product on the Hilbert space of states
\begin{equation}\label{eqn:fourierkernel}
\langle u^+ |u^- \rangle ~ = ~ \dfrac{1}{\sqrt{ 2\pi \lambda}} \, \exp\left( \frac{iu^+ u^-}{\lambda} \right) \, ,
\end{equation}
that expresses the Fourier transform kernel between the two bases. We may again realize the algebra if $\hat u^-$ acts on $ \langle u^+ |u^- \rangle $ and $\langle u^+ |x \rangle $ as $-i \lambda \partial_{u^+}$ while $\hat u^+$ acts on $\langle  u^-|u+\rangle $ and $ \langle u^-|x\rangle$ as $i \lambda \partial_{u^-}$. To endow the model with dynamics, we now turn to the Hamiltonian for each oscillator/partial wave
\begin{align}
H_{l m} ~ &= ~ \, \dfrac{1}{2} \left(p_{l m}^2 - x_{l m}^2\right) \nonumber \\
&= ~ \, \dfrac{1}{2}(u^+_{l m} u^-_{l m} + u^-_{l m} u^+_{l m}) \, ,
\end{align}
which may also be written as
\bea
H ~ = ~ \mp \, i \, \lambda  \left( u^\pm \partial_{u^\pm} + \dfrac{1}{2} \right) 
\eea
in the $u^\pm$ bases where we drop the $l , m$ indices. Physically the wave-function can be taken to correspond to a wave coming from the right which after scattering splits into a transmitted piece that moves on to the left and a reflected piece that returns to the right. The other wave function can be obtained from this one by a reflection $x \to -x$. The light-cone coordinates describe these left/right movers and simplify the description of scattering since the Schr\"{o}dinger equation becomes a first order partial differential equation. Moreover, the energy eigenfunctions are simply monomials of $u^\pm$ while in the $x$ representation the energy eigenfunctions are more complicated parabolic cylinder functions. In particular, for each partial wave the Schr\"{o}dinger equation in light-cone coordinates is:
\be
i \, \lambda \, \partial_t \psi_{\pm}\left(u^\pm , t\right) ~ = ~\mp  i \, \lambda \, \left(u^\pm \partial_{u^\pm} + 1/2 \right) \psi_{\pm}\left(u^\pm, t\right) \, 
\ee
with solutions
\be
\psi_{\pm}\left(u^\pm , t\right) ~ = ~ e^{\mp t/2} \, \psi_{\pm}^0 \left(e^{\mp t} u^{\pm}\right) \, .
\ee
This can also be written in bra/ket notation as:
\be
\langle u^\pm | \psi^\pm (t) \rangle ~ = ~ \langle u^\pm | e^{\frac{i}{\lambda} \hat H t} | \Psi^\pm_0 \rangle ~ = ~ e^{\mp \frac{t}{2}} \langle e^{\mp t} u^\pm | \Psi^\pm_0 \rangle \, .
\ee
The time evolution for the basis states is given by
\begin{align}
e^{\frac{i}{\lambda} Ht} |u^\pm \rangle ~ &= ~ e^{\pm{t\over 2}} |e^{\pm t} u^\pm \rangle \cr
\langle u^\pm| e^{\frac{i}{\lambda} Ht} ~ &= ~ e^{\mp {t\over 2}} \langle e^{\mp t} u^\pm | \cr
 \langle u^+| e^{\frac{i}{\lambda} H t}|u^- \rangle ~ &= ~ { 1 \over \sqrt{  2 \pi \lambda} } e^{-{t\over 2}}\exp\left( \frac{i}{\lambda} u^+ u^- e^{-t}\right)
\end{align}
\begin{figure}[t]
\centering
\includegraphics[width=70mm]{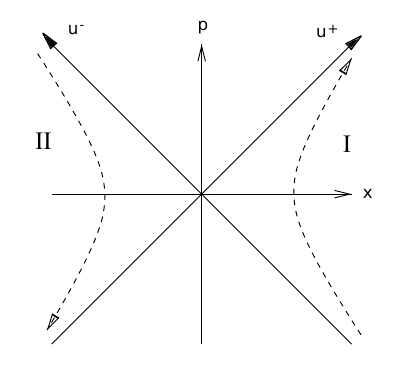}
\caption{The scattering diagram.}
\label{fig:scatteringdiagram}
\end{figure}
In the conventions of Figure \ref{fig:scatteringdiagram}, it is easy to see that ingoing states can be labelled by the $u^+$ axis while the outgoing ones by the $u^-$ axis. Since the potential is unbounded, the Hamiltonian has a continuous spectrum. In the $u^+$ representation the energy eigenstates with eigenvalue $\epsilon$ are
\begin{equation}
\dfrac{1}{\sqrt{2 \pi \lambda}}(u^+)^{i \frac{\epsilon}{\lambda} - \frac{1}{2}} \nonumber \, .
\end{equation}
The singularity at $u^+=0$ leads to a two fold doubling of the number of
states. This is understood to be arising from the existence of the two regions (I - II) in the scattering diagram. From now on we use  $|\epsilon,  \alpha^+ \rangle_{\text{in}}$ and $|\epsilon, \alpha^- \rangle_{\text{out}}$ for the in and outgoing energy eigenstates with the labels $\alpha^+ = \pm$ , $\alpha^- = \pm$ to denote the regions I and II. While we have four labels, we are still only describing waves in the two quadrants (I-II) with two of them for ingoing waves and two for outgoing ones. The in-states may be written as
\begin{align}
&\langle u^+|\epsilon, + \rangle_{\text{in}}=
\begin{cases}{1 \over \sqrt{2\pi \lambda}}{(u^+)}^{i\frac{\epsilon}{\lambda} - \half}& u^+>0\cr 0 & u^+<0\end{cases} \quad &\langle u^+|\epsilon,  - \rangle_{\text{in}} =
\begin{cases}0& u^+>0\cr {1 \over \sqrt{2\pi \lambda}}(-u^+)^{i\frac{\epsilon}{\lambda}-\half} & u^+<0\end{cases} \nonumber
\end{align}
describing left and right moving ingoing waves for the regions I and II respectively. Similarly, the natural out basis is written as
\begin{align}
 &\langle u^-|\epsilon, + \rangle_{\text{out}}=
 \begin{cases}{1 \over \sqrt{2\pi \lambda}}{(u^-)}^{-i\frac{\epsilon}{\lambda}-\half}& u^- >0\cr 0 & u^- <0\end{cases} \quad &\langle u^-|\epsilon, - \rangle_{\text{out}}=
 \begin{cases} 0& u->0\cr {1 \over \sqrt{2\pi \lambda}}(-u^-)^{-i\frac{\epsilon}{\lambda}-\half} & u^-<0 \end{cases} \nonumber
\end{align}
to describe the right and left moving outgoing waves for the regions I and II respectively. Therefore, time evolution of the energy eigenstates
\begin{equation}
\langle u^+|\epsilon, +  \rangle_{\text{in}} (t) ~ = ~ \dfrac{1}{\sqrt{2 \pi \lambda}} e^{- i \frac{\epsilon}{\lambda} t} (u^+)^{i \frac{\epsilon}{\lambda} -\half} ~ = ~ \dfrac{1}{\sqrt{2 \pi \lambda}} e^{-\frac{\rho^+}{2}} e^{- i \frac{\epsilon}{\lambda} t} e^{i \frac{\epsilon}{\lambda} \rho^+} 
\end{equation}
implies that they correspond to the Rindler relativistic plane-waves\footnote{Normalised in the $u^\pm$ basis.} moving with the speed of light in the tortoise-coordinates if we identify the quantum mechanical time with Rindler time $t=\tau$ and the inverted harmonic oscillator energy with the Rindler momentum via $\kappa \lambda = \epsilon$. This means that the energy of the eigenstates of the non-relativistic inverted oscillator, when multiplied by $\lambda$, can also be interpreted as the energy/momentum of the Rindler relativistic plane waves of the previous section. This allows us to write down any ingoing state in terms of these Rindler plane waves. As we have seen, the unitary operator relating the $u^\pm$ representations is given by the fourier kernel \eqref{eqn:fourierkernel} on the whole line that acts on a state as 
\be
\psi_{\text{out}} (u^-) ~ = ~ \left[ \hat{\mathcal S} \psi_{\text{in}} \right] (u^-)= \int_{-\infty}^\infty \dfrac{du^+}{\sqrt{2 \pi \lambda}} e^{-i u^+ u^- \over \lambda} \psi_{\text{in}}(u^+).
\ee
It is now clear that repeating the calculations of the previous section results in the same S-Matrix, rather trivially. However, to make the connection to the eigenstates of the inverted harmonic oscillator transparent, we will derive it in a more conventional manner. To represent the action of the kernel on energy eigenstates, we split it into a $2\times 2$ matrix that relates them as follows:
\be
&\begin{pmatrix} |\epsilon, + \rangle_{out} \cr |\epsilon, - \rangle_{out} \end{pmatrix}
=&\cal \hat S \begin{pmatrix}|\epsilon, + \rangle_{in} \cr |\epsilon, - \rangle_{in} \end{pmatrix}
 \ee
The fastest method to find each entry is to compute the in-going energy eigenstates in the out-going position basis and vice versa using the insertion of a complete set of states of the form 
\begin{equation}
\langle u^- | \epsilon \rangle_{\text{in}} ~ = ~ \int_{-\infty}^\infty du^+ \langle u^- | u^+ \rangle \langle u^+ | \epsilon \rangle_{\text{in}} \, .
\end{equation}
The results are
\begin{align}
\langle u^-|\epsilon, \pm \rangle_{\text{in}} ~ &= ~ \lambda^{i \epsilon \over \lambda} e^{\mp i \pi \over 4} e^{\pm \frac{\pi \epsilon}{2 \lambda}} \Gamma\left(\dfrac{1}{2} + i \dfrac{\epsilon}{\lambda} \right)
\dfrac{(\alpha^-| u^-|)^{-i\frac{\epsilon}{\lambda} - \half}}{\sqrt{2\pi \lambda}} \\
\langle u^+|\epsilon, \pm \rangle_{\text{out}} ~ &= ~  \lambda^{-i \epsilon \over \lambda} e^{\pm i \pi \over 4} e^{\pm\frac{\pi \epsilon}{2 \lambda}} \Gamma\left(\dfrac{1}{2} - i \dfrac{\epsilon}{\lambda} \right)
\dfrac{(\alpha^+ |u^+|)^{i\frac{\epsilon}{\lambda} - \half}}{\sqrt{2\pi \lambda}}  \, .
\end{align}
Each of these equations gives two results for each sign\footnote{For negative signs, one makes use of $(-1)^{i\epsilon/\lambda - 1/2}  = e^{- i \pi/2} e^{- \pi \epsilon/\lambda}$.} to yield:
\begin{align}
\mathcal{S} ~ &= ~ \dfrac{1}{\sqrt{2 \pi}} \exp\left(-i \, \dfrac{\epsilon}{\lambda} \, \log \lambda\right) \Gamma\left(\half - i  \dfrac{\epsilon}{\lambda}\right) \begin{pmatrix} e^{- i \frac{\pi}{4}} \, e^{- \frac{\pi \epsilon}{2 \lambda}} & e^{i \frac{\pi}{4}} \, e^{\frac{\pi \epsilon}{2 \lambda}} \\
e^{i \frac{\pi}{4}} \, e^{\frac{\pi \epsilon}{2 \lambda}} & e^{- i \frac{\pi}{4}} \, e^{- \frac{\pi \epsilon}{2 \lambda}} \end{pmatrix} \nonumber \\
&= ~ e^{i\Phi(\epsilon)} \, \exp\left(-i \, \dfrac{\epsilon}{\lambda} \, \log \lambda\right)
\begin{pmatrix}{ e^{ -i \pi/4}\over \sqrt{1 + e^{2\pi \epsilon/\lambda} }} & e^{  i \pi/4} \over \sqrt{1 + e^{-2\pi \epsilon/\lambda}}\\
e^{ i \pi/4} \over \sqrt{1 + e^{-2\pi \epsilon/\lambda}} &
e^{- i \pi/4} \over \sqrt{1 + e^{2\pi \epsilon/\lambda}}\end{pmatrix} \, , 
\end{align}
with the scattering phase $\Phi(\epsilon)$ being defined as
\begin{equation}
\Phi(\epsilon) ~ = ~ \sqrt{\dfrac{\Gamma\left(\frac{1}{2} - i  \frac{\epsilon}{\lambda} \right)}{\Gamma\left(\frac{1}{2} + i  \frac{\epsilon}{\lambda}\right)}} \, .
\end{equation}
Identifying parameters as $k_l \, \lambda_l = \epsilon_l$, we see that this precisely reproduces the S-Matrix derived in the previous section for every partial wave. In this model, it is clear that the competition between reflection and transmission coefficients is owed to the energy of the waves being scattered being larger than the tip of the inverted potential. \\

\subsection{A projective light-cone construction}
Although we had good reason to expect such an inverted harmonic oscillator realization of the black hole S-Matrix, there is, in fact, another way to derive it\textemdash using what is called a projective light-cone construction. This construction was first studied by Dirac and \cite{Rychkov:2016iqz,Weinberg:2010fx} provide a good modern introduction to the topic. The essential idea is to embed a null hyper-surface inside Minkowski space to study how linear Lorentz symmetries induce non-linearly realized conformal symmetries on a (Euclidean) section of the embedded surface. This allows us to relate the Rindler Hamiltonian -which can then be related directly to the Hamiltonian of the quantum mechanics model that describes the scattering- with the Dilatation operator on the horizon.  In a black hole background this construction is of course expected to hold only locally in the near horizon region. We first introduce $X=(x^\mu, x^{d-1}, x^{d} )$ with $\mu = 1, .., d-2$ (note that $\mu$ is a Euclidean index), where the light-cone coordinates are defined as $x^\pm = x^{d} \pm x^{d-1}$. Here, $x^{d}$ serves as the time coordinate \footnote{The null cone is described by the equation $X^2=0$ and a Euclidean section can be given as $x^+=f(x^\mu)$.}. The Minkowski metric $\eta_{MN}$ in these coordinates is given as
\begin{equation}
ds^2 ~ = ~  - dx^+ dx^- + dx_\mu dx^\mu \, ,
\end{equation}
which has an $SO(d-1,1)$ Lorentz symmetry. There is an isomorphism between the corresponding Lorentz algebra and the Euclidean conformal algebra in $d-2$ dimensions. To state this isomorphism, we first label the $d-2$-dimensional Euclidean conformal group generators as:
\begin{align}
P_\mu ~ &= ~ i\partial_\mu \quad &&\text{corresponding to translations,} \nonumber \\ 
M_{\mu\nu} ~ &= ~ i \left(x_\mu\partial_\nu-x_\nu\partial_\mu\right) \quad &&\text{to rotations,} \nonumber \\ 
D ~ &= ~ i x^\mu\partial_\mu \quad &&\text{to dilatations, and} \nonumber \\
K_\mu ~ &= ~ i \left(2x_\mu \left(x^\nu\partial_\nu\right) - x^2\partial_\mu\right) \quad &&\text{to special conformal transformations} \, .
\end{align}
The identification is now given as follows:
\begin{align}
J_{\mu\nu} ~ = ~ M_{\mu\nu} \, , \quad J_{\mu+} ~ = ~ P_\mu \, , \quad J_{\mu-} ~ = ~ K_\mu \, , \quad J_{+-} ~ = ~ D \, , 
\end{align}
where the $SO(3,1)$ Lorentz generators $J_{MN}$ are given by
\begin{equation}
J_{M N} ~ = ~ x_M p_N - x_N p_M \, .
\end{equation}
These satisfy the $SO(3, 1)\backsimeq SL(2,\mathbb{C})$ algebra. In particular the Dilatation operator on the two dimensional horizon is
\begin{equation}
D ~ = ~ J_{+ -} ~ = ~ x_+ p_- - x_- p_+ ~ = ~ \dfrac{1}{\lambda} \left(u^+ u^- + u^- u^+ \right) ~ = ~ \dfrac{1}{\lambda} H \, , 
\end{equation}
where in the second equality we used $u^\pm = x_\pm$ to connect to the light-cone coordinates of the previous sub-section and in the third equality, we made use of the back-reaction relations \eqref{eqn:partialwavebackreaction}. Interestingly enough, we see that an appropriately scaled Dilatation operator together with the back-reaction relations gives us exactly the Hamiltonian of the inverted oscillator. The scaling is also neatly realized in the relation between the quantum mechanical energy $\epsilon$ and the Rindler energy $\kappa$ to relate the two S-Matrices. \\

This construction via the light-cone projection could possibly shed more light on the relation between the black hole S-Matrix and string theoretic amplitudes. In the early papers on black hole scattering \cite{'tHooft:1991bd,'tHooft:1992zk,'tHooft:1996tq}, a striking similarity between the S-Matrix and stringy amplitudes was observed. The role of the string worldsheet was attributed to the horizon itself. It was noted that the string tension was imaginary. In the construction above, we found that the induced conformal symmetry on the horizon is Euclidean and that the Dilatation operator is mapped to the time-evolution operator (Rindler Hamiltonian) of the 4d Lorentzian theory. This led us to the unstable potential of the inverse harmonic oscillator. It may well be that the apparently misplaced factors of $i$ in the string tension is owed to the Euclidean nature of conformal algebra on the horizon. It would also be interesting to understand the role of possible infinite-dimensional local symmetries on the horizon/worldsheet \cite{Hawking:2015qqa,Hawking:2016msc} from the point of view of the quantum mechanics model, elaborating on the null cone construction. We leave this study to future work. \\

While the model is seemingly very simple, this is not the first time that such a model has been considered to be relevant for black hole physics \cite{Friess:2004tq,Maldacena2005}. However, previous considerations have found that these models do not correspond to 2d black hole formation owing to an insufficient density of states in the spectrum. Refining these considerations with the intuition that each oscillator as considered in this section corresponds to a partial wave of a 4d black hole, we find that our model may indeed be directly related to 4d black holes formed by physically collapsing matter. We provide evidence for this in Section \ref{sec:collapse}. In order to move on to which, however, it will be very useful for us to review the 2d string theory considerations of the past; this is what we now turn to.

\subsection{Relation to matrix models and 2-d string theory}
Hermitian Matrix Quantum Mechanics (MQM, henceforth) in the inverted harmonic oscillator was studied in connection with $c=1$ Matrix models and string theory in two dimensions. For more details, we refer the reader to \cite{Klebanov:1991qa,Martinec:2004td}. Here, we briefly review these results in order to point out various similarities and differences with our work. The Lagrangian of MQM is of the form
\begin{equation}
L ~ = ~ \dfrac{1}{2} \Tr \left[ \left(D_t M\right)^2 + M^2 \right] \qquad \text{with} \qquad D_t ~ = ~ \partial_t - i A_t \, ,
\end{equation}
where $A_t$ is a non-dynamical gauge field. The $N \times N$ Hermitian Matrices transform under $U(N)$ as $M \rightarrow U^\dagger M U$. The role of the non-dynamical gauge field is to project out the non-singlet states in the path integral.  Diagonalization of the matrices results in a Vandermonde factor in the path integral measure:
\be 
\mathcal{D} M ~ = ~ \mathcal{D} U \, \prod_i \, d x_i \, \prod_{i<j} (x_i - x_j)^2 \, .
\ee
This indicates a natural fermionic redefinition of the wave-functions into Slater determinants (in a first quantised description). The Hamiltonian of the system is, therefore, in terms of $N$ free fermions:
\be
\hat H \, \tilde{\Psi} ~ = ~ - \left(\dfrac{\hbar^2}{2} \sum_{i=1}^N \partial_{x_i}^2 + \dfrac{1}{2} x_i^2 \right) \, \tilde{\Psi} \qquad \text{with} \qquad \tilde{\Psi}(x_i)= \prod_{i<j} (x_i - x_j) \Psi(x_i) \, ,
\ee
with $\tilde{\Psi}(x_i)$ being the redefined fermionic wave-functions. Filling up the `Fermi-sea' up to a level $\mu$, allows for a definition of the vacuum. Clearly, all fermions are subject to the same chemical potential $\mu$ that is typically considered to be below the tip of the inverted oscillator. A smooth string world-sheet was argued to be produced out of these matrices in a double-scaling limit $\mu \rightarrow 0 , \hbar \rightarrow 0$ with a fixed inverse string-coupling defined by the ratio $\mu/\hbar \sim 1/g_s $. In this double-scaling limit, this theory describes string theory on a 2d linear dilaton background with coordinates described by time $t$ and the Liouville field $\phi$. The matrix model/harmonic oscillator coordinate $x$ is conjugate to the target space Liouville field via a non-local integral transformation \cite{Moore1992a}. In contrast to this picture, owing to a one-one correspondence between the 2d harmonic oscillators and 4d partial waves in our model, this integral transform is unnecessary. However, it has been argued in string theory that only the quadratic tip is relevant in this double-scaling limit, even in the presence of a generic inverted potential, emphasizing the universality of the quadratic tip. Whilst we do not have a similar stringy argument, we expect the ubiquitous presence of the quadratic potential to persist in our construction owing to the ubiquitous presence of the Rindler horizon in physical black holes formed from collapsing matter. A modern discourse with emphasis on the target space interpretation of the matrix model as the effective action of $N$ $D0$ branes may be found in \cite{McGreevy2004}. A natural second quantized string field theory description of the system where the fermionic wave-functions are promoted to fermionic fields may be found in \cite{Ginsparg1992,Klebanov:1991qa,Nakayama2004} and references therein. A satisfactory picture of free fermionic scattering in the matrix model was given in \cite{Moore1992b} via the following S-Matrix relation:
\begin{equation}
\hat S ~ = ~ i_{b \rightarrow f} \circ \hat S_{ff} \circ  i_{f \rightarrow b} 
\end{equation}
where even though the asymptotic tachyonic states are bosonic, one is instructed to first fermionize, then scatter the fermions in the inverted quadratic potential and then to bosonize again. The total S-matrix is unitary if the fermionic scattering is unitary and the bosonization spans all possible states. The logic of this expression resembles that of 't Hooft's S-matrix, where one first expands a generic asymptotic state into partial waves, expresses them in terms of near horizon Rindler parameters, scatters them with the given S-matrix that is similar to the one of 2d string theory before transforming back to the original asymptotic coordinates. At the level of the discussion now, it may already be noted that one important difference between the 2d string-theoretic interpretation of the matrix model and our 4d partial wave one is the nature of the transformations that relate asymptotic states to the eigenstates of the inverted harmonic oscillator. Additionally, and perhaps more importantly, in our construction, we have an entire collection of such harmonic oscillators/matrix models parametrised by $l, m$ that conspire to make up a 4d black hole. We present concrete evidence for this by studying time-delays and degeneracy of states in Section \ref{sec:collapse}. There are further differences between the 2d string theories and our construction, in order to present which, we need to proceed to a study of the spectrum of states in our model; this enables us to study growth of states in the two models. Finally, we also comment on a possible second quantization and appropriate MQM interpretation of our model in Section \ref{sec:discussion}.

\section{Combining the oscillators (partial waves)}\label{sec:collapse}
On the side of the macroscopic black hole in Section \ref{sec:macroSMatrix}, the calculation was done in an approximation where there is a pre-existing black hole into which degrees of freedom are thrown (as positions and momenta). It was then evident that the information that was sent into the black hole is completely recovered since the S-Matrix was unitary. Furthermore, the back-reaction computation told us exactly how this information is retrieved: in-going positions as out-going momenta and in-going momenta as out-going positions. However, a critical standpoint one may take with good reason would be to say that this is not good enough to tell us if a physical collapse of a black hole and complete evaporation of it is a unitary process. The calculation has not modeled a collapsing problem.\\

The picture to have in a realistic collapse is that of an initial state that evolves in time to collapse into an intermediate black hole state which then subsequently evaporates to result in a final state that is related to the initial one by a unitary transformation. Naturally, the corresponding macroscopic picture is that of a strongly time-dependent metric. Heuristically, one may think of the total S-Matrix of this process as being split as
\begin{equation}
\hat S ~ = ~ \hat S_{\mathscr{I}^- \rightarrow \text{hor}^-} ~ \hat S_{\text{hor}^- \rightarrow \text{hor}^+} ~ \hat S_{\text{hor}^+ \rightarrow \mathscr{I}^+ }
\end{equation}
where $\hat S_{\mathscr{I}^- \rightarrow \text{hor}^-}$ corresponds to evolution from asymptotic past to a (loosely defined) point in time when gravitational interactions are strong enough for the collapse to begin, $\hat S_{\text{hor}^- \rightarrow \text{hor}^+}$ to the piece that captures all the `action'\textemdash insofar as collapse and evaporation are concerned\textemdash take place and finally $\hat S_{\text{hor}^+ \rightarrow \mathscr{I}^+ }$ represents the evolution of the evaporated states to future infinity. The horizon\textemdash being a teleological construction that can be defined only if one knows the global structure of spacetime\textemdash has a time dependent size and location in a collapse/evaporation scenario but for us will nevertheless comprise the locus of spacetime points where the backreaction effects are important. Therefore, we use subscripts $\text{hor}^\pm$ to refer to it, at different points in time, in the above heuristic split. Thought of the total evolution this way, it is clear that the most important contribution arises from the part of the matrix that refers to the region in space-time where gravitational back-reaction cannot be ignored. The other pieces are fairly well-approximated by quantum field theory on an approximately fixed background. Nevertheless, in the intermediate stage, the metric is inherently time-dependent. \\

At the outset, let it be stated that we will not get as far as being able to derive this metric from the quantum mechanics model. We may ask if there are generic features of the black hole that we have come to learn from semi-classical analyses that can also be seen in this model. We will focus on two important qualitative aspects of (semi-classical) black holes:
\paragraph{Time-delay} A physical black hole is not expected to instantaneously radiate information that has been thrown into it. There is a time-delay between the time at which radiation begins to be received by a distant observer and the time at which one may actually recover in-going information. In particular, given an in-state that collapses into a black hole, we expect that the time-scale associated to the scattering process is `long'. In previous studies of 2d non-critical string theory, it was found that with a single inverted harmonic oscillator, the associated time-delay is not long enough to have formed a black hole \cite{Schoutens:1993hu,Karczmarek:2004bw,Friess:2004tq}. However, with the recognition that each oscillator corresponds to a partial wave and that a collection of oscillators represents a 4d black hole, we see that the black hole degeneracy of states arises from the entire collection while the time-delay associated to each oscillator is the time spent by an in-going mode in the scattering region; the latter being more reminiscent of what one might call `scrambling time'.
\paragraph{Approximate thermality} As Hawking famously showed \cite{Hawking:1974sw}, the spectrum of radiation looks largely thermal for a wide range of energies. One way to probe this feature is via the number operator\textemdash which, for a finite temperature system, can be written as $\bra \hat N (\omega)\ket= \rho(\omega) f(\omega)$ with $\rho(\omega)$ being the density of states and $f(\omega)$ the appropriate thermal distribution for Fermi/Bose statistics. Given that the S-Matrix is unitary, we know that this notion of temperature and thermality of the spectrum is only approximate. Notwithstanding this, a detector at future infinity should register this approximately thermal distribution for a large frequency range. \\

In what follows, we will study whether the S-Matrix corresponding to our collection of oscillators in the model presented in Section \ref{sec:micromodel} displays both these properties.

\subsection{Time delays and degeneracy of states}
We have seen that the total scattering matrix associated to four-dimensional gravity can be seen as arising via a collection of inverted harmonic oscillators, each with a different algebra differentiated by $\lambda_l$ in, say, \eqref{eqn:macroalgebra}. One canonical way to study life-times in scattering problems in quantum mechanics is via the time-delay matrix, which is defined as:
\begin{equation}\label{eqn:timedelay}
\Delta t_{ab} ~ = ~ \Re\left(i \, \sum_c \, S^\dagger(k_l,\lambda_l)_{ac} \left(\dfrac{d S(k_l,\lambda_l)}{d k}\right)_{cb}\right) \, .
\end{equation}
Each matrix element above encodes the time spent by a wave of energy $k_l$ in the scattering region in the corresponding channel. The trace of this matrix, called Wigner's time delay $\tau_l$, captures the total characteristic time-scale associated to the entire scattering process. Said another way, should we start with a generic in-state that undergoes scattering and is then retrieved in the asymptotic future as some out-state, the trace of the above matrix associates a life-time to the intermediate state \cite{deBianchi,deCarvalho200283}. For large energies $k_l$, using $S\left(k_l,\lambda_l\right)$ in \eqref{eqn:SMatrix2}, the Wigner time-delay associated to the scattering of a single oscillator can be calculated to scale as $\tau \sim \log\left(\lambda_l \, k_l\right)$. This is the same result as was found in the 2d string theory literature \cite{Schoutens:1993hu,Karczmarek:2004bw,Friess:2004tq} and was argued to not be long-enough for black hole formation. Based on these black hole non-formation results in the matrix quantum mechanics, it was suggested that studying the non-singlet sectors would shed light on 2d black hole formation\cite{Kazakov2002,Banks2015}. Despite some efforts in relating the adjoint representations with long-string states \cite{Maldacena2005}, a satisfactory Lorentzian description is still missing. Anticipating our result prematurely, our model does not suffer from these difficulties as it is to describe a 4d black hole with a collection of oscillators. Merely the $s$-wave oscillator in our model would mimic the singlet sector in matrix quantum mechanics\footnote{It would be very interesting if higher $l$ modes can be described as non-singlets of a matrix model.}.\\

The above time delay $\tau$ may also be interpreted as a density of states associated to the system. The inverted potential under consideration implies a continuous spectrum. In order to discretize which, to derive the density of states, the system must be stabilized\textemdash by putting it in a box of size $\Lambda$, for instance. Demanding that the wavefunctions vanish at the wall and regulating the result by subtracting any cut-off dependent quantities, the density of states may be computed from the scattering phase $\Phi$ defined via $\mathcal{S}\left(k_l,\lambda_l\right) = \exp\left[i \, \Phi\left(k_l,\lambda_l\right)\right]$ as $\rho(\epsilon_l)= d \Phi/d \epsilon_l$ \cite{Kazakov:1990ue}. The result is exactly the same as what we get from computing the time delay using the scattering matrix \eqref{eqn:SMatrix2} and the time-delay equation \eqref{eqn:timedelay} to find a Di-Gamma function $\psi^{(0)}$
\begin{align}\label{eqn:densityofstates}
\rho\left(\epsilon_l\right) ~ = ~ \tau_l ~ &= ~ \, \dfrac{2}{\lambda_l} \Re \left[\psi ^{(0)}\left(\dfrac{1}{2} - i \dfrac{\epsilon_l}{\lambda_l}\right) + \log\left(\lambda_l\right)\right] \nonumber \\
&= ~ \Re \left[\sum_{n=0}^\infty ~ \dfrac{2}{i \epsilon_l - \lambda_l \left(n+\frac{1}{2}\right)} + \dfrac{2}{\lambda_l} \log\left(\lambda_l\right) \right] \, ,
\end{align}
where we subtracted an infinite energy independent constant and the factor of $i$ is because we are describing an unstable system. This density of states may be used to define formally a partition function for each partial wave (with Hamiltonian $\hat{H}_{l m}$), where the energy eigenstates contributing to the partition function will have been picked out by the poles of the density $\rho\left(\epsilon_l\right)$. However, in our model, we see that there are many oscillators in question. Should we start with an in-state made of a collection of all oscillators instead of a single partial wave, we may first write down the total S-Matrix as a product of the individual oscillators as
\begin{equation}\label{eqn:totSMatrix}
\mathcal{S}_{\text{tot}} ~ = ~ \prod_{l=0}^\infty \, \mathcal {S}\left(k_l, \lambda_l\right) \, ,
\end{equation}
assuming that different partial waves do not interact. One may correct for this by adding interaction terms between different oscillators. To compute the time-delay associated to a scattering of some in-state specified by a given total energy involves an appropriately defined Wigner time-delay matrix as
\begin{equation}\label{eqn:totTime}
\tau_{\text{tot}} ~ = ~ \text{Tr} \left[\Re\left(- i \left(\mathcal{S}^\dagger_{\text{tot}}\right)_{ac} \left(\dfrac{d \mathcal{S}_{\text{tot}}}{dE_{\text{tot}}}\right)_{cb}\right)\right] \, ,
\end{equation}
where this equation makes sense only if we have defined a common time evolution and unit of energy for the total system/collection of partial waves. We will elaborate on this in a while.
Now, even in the spherically symmetric approximation, to write the total S-Matrix as a function of merely one coarse-grained energy $E_{\text{tot}}$ is not a uniquely defined procedure. However, our intuition that each partial wave may be thought of as a single-particle oscillator allows us to compute the density of states in a combinatorial fashion. We will see that the degeneracy of states associated to an intermediate long-lived thermal state arises from the various ways in which one might distribute a given total energy among the many available oscillators. Given a total energy $E_{\text{tot}}$, we now have the freedom to describe many states, each with a different distribution of energies into the various available oscillators. From the poles in the density defined in \eqref{eqn:densityofstates}, we see that each oscillator has energies quantized as\footnote{The seemingly disconcerting factor of $i$ is just owed to the fact that we have scattering states as opposed to bound ones.} 
\begin{align}
\epsilon_l ~ &= ~ i \lambda_l \left(n_l + \dfrac{1}{2}\right) \, .
\end{align}
This allows us to measure energies in units of $c$, where $c$ is defined implicitly via $\lambda_l \left(l^2 + l + 1\right) = c$. Therefore, in these units, the energies are `quantized' as
\begin{equation}
\dfrac{\epsilon_l}{i \, c} ~ = ~ \dfrac{1}{l^2 + l + 1} \left(n_l + \dfrac{1}{2}\right) \, .
\end{equation}
Now, given some total energy $E_{\text{tot}}$, we see that any oscillator may be populated with a single particle state carrying energy such that $n_l = E_{\text{tot}} \left(l^2 + l + 1\right)$, where we leave out the half integer piece for simplicity. Importantly, we see that there exist `special' states coming from very large $l$-modes even for very small energies. For example, an energy of $1$ could arise from a very large $l$-mode with the excitation given by $n_l = \left(l^2 + l + 1\right)$. This is rather unsatisfactory for one expects that it costs a lot of energy to create such states. Moreover, there is an interplay between the log term in the growth of states and the behaviour of the di-gamma function that we are unable to satisfactorily take into account. There is an additional problem which is that the energy of each partial wave is measured in different units that are $l$ dependent; this means that they also evolve with different times. We thus conclude that this is not the correct way to combine the different oscillators. \\
  
There is a rather beautiful way to resolve all these three problems via a simple change of variables that we turn to next. It will allow us to interpret the above cost of energy as relative shifts of energies with respect to a common ground state. Additionally these relative shifts also cure the above interplay; there will simply be no log term in the density of states. Finally, this will also introduce a canonical time evolution for the entire system, resulting in one common unit of energy.

\subsection{Exponential degeneracy for the collection of oscillators}
In order to combine the different oscillators and define a Hamiltonian for the total system we need to get rid of the $l$ dependence in the units of energy used for different oscillators. It turns out that this is possible  by rewriting the black hole algebra. Moreover using these new variables, the relation between 't Hooft's black hole S-Matrix for an individual partial wave and the one of 2d string theory of type II \cite{Moore1992b} can be made manifest. To make this connection transparent, we again start with a collection of inverse harmonic oscillators and the following Hamiltonian for the total system
\begin{align}
H_{tot} ~ &= ~ \sum_{l,m} \, \dfrac{1}{2} \left(\tilde{p}_{l m}^2 - \tilde{x}_{l m}^2\right) \nonumber \\
&= ~ \sum_{l,m} \, \dfrac{1}{2} \left(\tilde{u}^+_{l m}  \tilde{u}^-_{l m} + \tilde{u}^-_{l m} \tilde{u}^+_{l m}\right) \, ,
\end{align}
but this time imposing the usual $\lambda$-independent commutation relations $[\tilde{u}^+_{l m} \tilde{u}^-_{l' m'}]= i \delta_{l l'} \delta_{m m'}$. The $\lambda$ dependence will come through via an assignment of a chemical potential $\mu(\lambda)$ for each oscillator; this assignment is to be thought of as a different vacuum energy for each partial wave. Following \cite{Moore1992b}, one may then derive an S-Matrix for this theory. To match this to the one of 't Hooft for any given partial wave, one must identify the chemical potential and energy parameters as $\mu =1 / \lambda$ and the Rindler energy $k=\omega+ \mu  =\omega + 1 / \lambda$. It is worth noting that in the reference cited above, only energies below the tip of the inverted potential were considered, resulting in a dominant reflection coefficient. In contrast 't Hooft's partial waves carry energies higher than the one set by the tip of the potential. Consequently, to make an appropriate identification of 2d string theory with the partial wave S-Matrix, an interchanging of the reflection and transmission coefficients is necessary. From a matrix quantum mechanics point of view, it may additionally be noted that the partial wave parameter $\lambda$ may be absorbed in either the Planck's constant or the chemical potential to leave the string coupling of each partial wave fixed as $g_s \sim \hbar/\mu \sim c/(l^2+l+1)$. This indicates that as we increase the size of the black hole or we consider higher $l$ partial waves the corresponding string coupling becomes perturbatively small. \\
Writing out the energies of the various partial waves with the above identification, we have
\begin{equation}
k_l ~ = ~ \omega_l + \frac{l^2+l+1}{c} \, , \qquad \text{and} \qquad  E^{\text{Rindler}}_{\text{tot}} ~ = ~ \sum_l k_l \, .
\end{equation}
At this stage, the labels $\omega_l$ are continuous energies. However, discretizing the spectrum as before, by putting the system in a box, we arrive at discrete energies\footnote{Note again the relative factor of $i$ that indicates that the harmonic oscillator levels have to do with decaying/scattering states while $l$'s are naturally discrete due to the $S^2$ being compact.} 
\begin{equation}\label{eqn:totenergy}
c \, E_{\text{tot}} ~ = ~ \sum_l \, \left[ i \, c \,  \left( n_l + \half \right) + l^2 + l + 1\right] \, ,
\end{equation}
for every individual oscillator. Without a detour into this 2d string theory literature, we may have alternatively arrived at this spectrum from the quantum mechanics model in \ref{sec:micromodel} via the following identifications:
\begin{equation}
\epsilon_l ~ \longrightarrow ~ 1 + \lambda_l \, \omega_l \qquad \text{and} \qquad \lambda_l ~ \longrightarrow ~ \dfrac{1}{\mu_l} \, .
\end{equation}
\begin{figure}
\centering
\vspace{-10pt}
\includegraphics[width=0.9\textwidth]{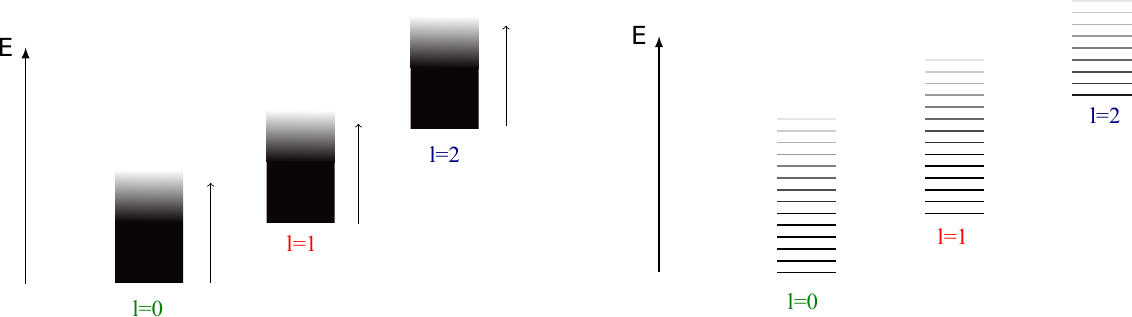}
\caption{Spectrum of the collection of oscillators. On the left we depict the original continuous spectrum, while on the right the discrete spectrum with our choice of regularising the potential (cutoff).}
\label{fig:levels}
\end{figure} 
While the model presented in Section \ref{sec:micromodel} makes the algebra manifest, the above identification of parameters to relate to the model with a $\lambda$-independent algebra makes the physical interpretation of the relative shifts in energies between the partial waves manifest and allows for a consistent definition of time and energy for the total system. \\

This allows us to rewrite our S-Matrix $S\left(\epsilon_l/\lambda_l\right)$ as a function of two variables $\omega_l$ and $\mu_l$ as $S\left(\omega_l,\mu_l\right)$. With this change of variables, we recover exactly the S-Matrix of the 2d matrix models discussed in the literature (for each partial wave) and moreover, it is now natural to interpret $\mu$ as a chemical potential that defines the vacuum energy of each partial wave on top of which we can further excite the various scattering modes. However, since $\mu = \mu_l$ is now $l$ dependent in our collection of oscillators, it gives us a natural way to interpret how the combined system behaves. To excite a very large $l$ oscillator, one first has to provide sufficient energy that is equal to $\mu_l \sim \left(l^2 + l + 1\right)$. Therefore, we naturally see that exciting a large $l$-oscillator costs energy! The physical spectrum may be depicted as in Figure \ref{fig:levels}, where we have chosen an arbitrary value for the ground-state energy with $E = 0$. each oscillator labelled by $l$ and excitations above them by $n_l$. The various oscillators are shifted by the chemical potential $\mu_l$. And the vacuum is defined to be the one with all Rindler energies $k_l$ set to zero. Now, given an initial state carrying a total energy of $E_{\text{tot}}$, we are left with a degeneracy of states that may be formed by distributing this energy among the many available oscillators. The larger this energy, the more oscillators we may distribute it into and hence the larger the degeneracy. The degeneracy associated to equation \eqref{eqn:totenergy}, without the chemical potential shift, is merely asking for the number of sets of all integers $\{n_l\}$ that add up to $E_{\text{tot}}$. These are the celebrated partitions into integers that\textemdash as Ramanujan showed\textemdash grow exponentially. Clearly, for large total energy, our degeneracy grows similarly at leading order. However, the chemical potential shift slows down the growth polynomially compared to the partitioning into integers owing to the fact that for a given $E_{\text{tot}}$, only approximately $\sqrt{E_{\text{tot}}}$ number of oscillators are available. It is worthwhile to note that, in this simplistic analysis, we have ignored the degeneracy arising from the $m$ quantum number; accounting for which clearly increases the growth of states. Therefore, we find that the model exhibits an exponential growth in its density of states! This shares striking resemblance to the Hagedorn growth of density of states in black holes and is an indication that we have captured the associated microstates of the spherically symmetric Schwarzschild solution.\\

As a conservative estimate, we may start with some total energy $E_{\text{tot}}$ and a fixed set of oscillators that are allowed to contribute to it. This allows us to sum over the contribution arising from the $\left(l^2 + l + 1\right)c^{-1}$ piece in \eqref{eqn:totenergy} to be left with some subtracted total energy $\tilde{E}_{\text{tot}}$ that is to be distributed among the $n_l$ excitations over each of the available oscillators. Clearly, this grows exponentially much as the partitions into integers does, with the subtracted energy $\tilde{E}_{\text{tot}}$. This is given by the famous Hardy-Ramanujan formula for the growth of partitions of integers:
\begin{equation}
p\left(n\right) ~ \sim ~ \exp\left(\pi \sqrt{\dfrac{2 \, n}{3}}\right) \, .
\end{equation}
Identifying $n$ with the integer part of $\tilde{E}_{\text{tot}}$, we see the desired exponential growth. And considering that the same total energy may be gained from choosing different sets of oscillators to start with, increases this degeneracy further, in equal measure. While imposing the antipodal identification of 't Hooft\textemdash which we discuss in Section \ref{sec:discussion}\textemdash reduces this degeneracy, the exponential growth of states remains. How one may derive the Schwarzschild entropy from this degeneracy requires a truly microscopic understanding of the parameter $\lambda$. We suggest a way forward towards the end of this article but leave a careful study to future work.

\section{Discussion}\label{sec:discussion}
In this article, we have constructed a quantum mechanics model that reproduces 't Hooft's black hole S-Matrix for every partial wave, using which, we provided non-trivial evidence that it corresponds to a black hole S-Matrix; owing to the appropriate exponential density of states. Several questions, though, remain unanswered. The only degrees of freedom in question were the momenta and positions of ingoing modes. One may add various standard model charges, spin, etc. to see how information may be retrieved by the asymptotic observer. \\

Dynamically speaking, gravitational evolution is expected to be very complicated in real-world scenarios. We have merely approximated it to one in which different spherical harmonics do not interact. While incorporating these interactions may be very difficult to describe in gravity, they are rather straightforward to implement in the quantum mechanical model; one merely introduces interaction terms coupling different oscillators. Exactly what the nature of these interactions is, is still left open. \\

The complete dynamics of the black hole includes a change in mass of the black hole during the scattering process. In this article, we chose to work in an adiabatic approximation, where this is ignored. The corresponding approximation in the inverted oscillators is that the potential is not affected by the scattering waves. This would then mean that in a more realistic setting, the quadratic potential changes due to the waves that scatter off it. The change in the form of the inverted potential due to a scattering mode can be calculated \cite{Asplund:2010xh}. We hope to work on this in the future and we think that this gives us a natural way to incorporate the changes in the mass of the black hole. Another possible avenue for future work is to realise a truly microscopic description of the S-matrix, either in the form of a matrix model or a non-local spin model having a finite-dimensional Hilbert space from the outset, where the inverse harmonic potential or emergent $SL(2,\mathbb{R})$ symmetries are expected to arise after an averaging over the interactions between the microscopic degrees of freedom. Some models with these properties can be found in \cite{Kitaev,Anninos:2014ffa,Anninos:2015eji,Maldacena:2016hyu}.

\begin{figure}[t]
\centering
\includegraphics[width=0.7 \textwidth]{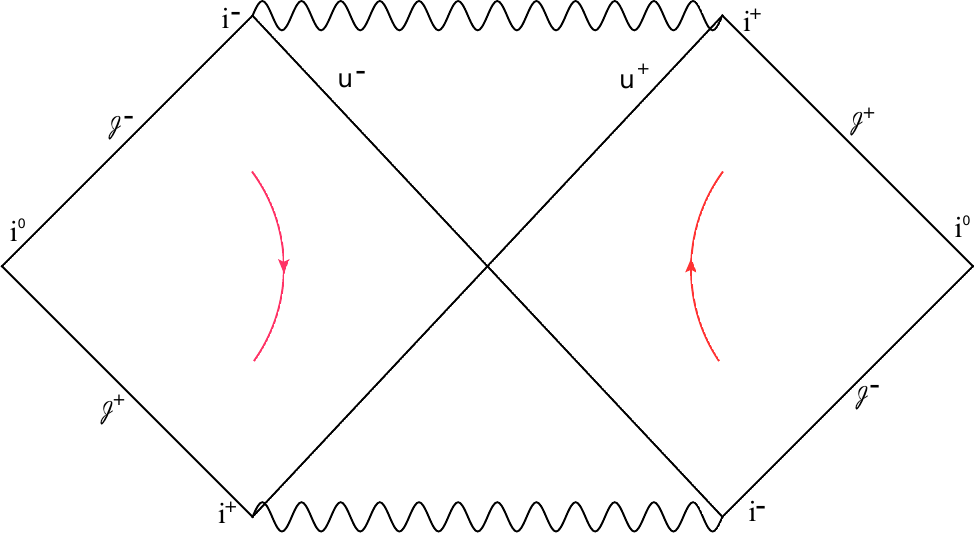}
\caption{The Penrose diagram. Each point in the conformal diagram originally corresponds to a different sphere. After antipodal identification, the points $(u^+, u^-)$ and $(-u^+, -u^-)$ correspond to antipodal points on a common sphere as given in \eqref{eqn:antipodalidentification}. The red lines indicate the arrows of time.}
\label{fig:2sided}
\end{figure}

\paragraph{Antipodal entanglement}
Unitarity of the S-Matrix demands that both the left and right exteriors in the two-sided Penrose diagram need to be accounted for; they capture the transmitted and reflected pieces of the wave-function, respectively. In the quantum mechanics model, there appears to be an ambiguity of how to associate the two regions I and II of the scattering diagram in Fig. \ref{fig:scatteringdiagram} to the two exteriors of the Penrose diagram. We saw, in the previous section, that the quantum mechanical model appears to support the creation of physical black holes by exciting appropriate oscillators. Therefore, in this picture there is necessarily only one physical exterior. To resolve the issue of two exteriors, it was proposed that one must make an antipodal identification on the Penrose diagram \cite{Hooft:2016itl}; see figure \ref{fig:2sided}. Unitarity is arguably a better physical consistency condition than a demand of the maximal analytic extension. The precise identification is given by $x \rightarrow J x$ with\footnote{Note that the simpler mapping of identifying points in $I , II$ via  $(u^+, u^-, \theta , \phi) \leftrightarrow (-u^+, -u^-, \theta, \phi)$ is singular on the axis $u^+, u^- =0$.}
\begin{equation}\label{eqn:antipodalidentification}
J: \quad (u^+ , u^- , \theta , \phi) ~\longleftrightarrow ~ \left(- u^+ , - u^- ,\pi - \theta, \pi + \phi\right) \, .
\end{equation}
Note that $J$ has no fixed points and is also an involution, in that $J^2=1$. Such an identification implies that spheres on antipodal points in the Penrose diagram are identified with each other. In particular, this means 
\begin{equation}
u^\pm \left(\theta, \phi\right) ~ = ~ - u^\pm \left(\pi - \theta, \pi + \phi\right) \qquad \text{and} \qquad p^\pm \left(\theta, \phi\right) ~ = ~ - p^\pm \left(\pi - \theta, \pi + \phi\right) \, .
\end{equation}
Therefore, noting that the spherical harmonics then obey $Y_{l,m} \left(\pi - \theta, \pi + \phi\right) = \left(-1\right)^l Y_{l,m} \left(\theta,\phi\right)$, we see that only those modes with an $l$ that is odd contribute. However, owing to the validity of the S-Matrix only in the region of space-time that is near the horizon, this identification is presumably valid only in this region. Global identifications of the two exteriors have been considered in the past \cite{Gibbons:1986dd,Sanchez:1986qn,Parikh:2002py}. The physics of the scattering, with this identification is now clear. In-going wave-packets move towards the horizon where gravitational back-reaction is strongest according to an asymptotic observer. Most of the information then passes through the antipodal region and a small fraction is reflected back. Turning on quantum mechanics implies that ingoing position is imprinted on outgoing momenta and consequently, an highly localised ingoing wave-packet transforms into two outgoing pieces\textemdash transmitted and reflected ones\textemdash but both having highly localised momenta. Their positions, however, are highly de-localised. This is how large wavelength Hawking particles are produced out of short wavelength wave-packets and an IR-UV connection seems to be at play. Interestingly, the maximal entanglement between the antipodal out-going modes suggests a wormhole connecting each pair \cite{Maldacena:2013xja}; the geometric wormhole connects the reflected and transmitted Hilbert spaces. Furthermore, as the study of the Wigner time-delay showed, the reflected and transmitted pieces arrive with a time-delay that scales logarithmically in the energy of the in-going wave. This behaviour appears to be very closely related to scrambling time (not the lifetime of the black hole) and we leave a more detailed investigation of this feature to the future. One may also wonder why transmitted pieces dominate the reflected ones. It may be that the attractive nature of gravity is the actor behind the scene.
 
\paragraph{Approximate thermality}
We now turn to the issue of thermality of the radiated spectrum. Given a number density, say $N^{\text{in}}(k)$ as a function of the energy $k$, we know that there is a unitary matrix that relates it to radiated spectrum. This unitary matrix is precisely the S-Matrix of the theory. The relation between the in and out spectra is given by $N^{\text{out}}(k)=S^{\dagger}N^{\text{in}}(k)S$. Using the explicit expression for the S-Matrix \eqref{eqn:SMatrix2}, we find
\begin{align}
N_{++}^{\text{out}}(k)~ &= ~ \frac{N_{++}^{\text{in}}(k)}{1+e^{2\pi k}}+\frac{N_{--}^{\text{in}}(k)}{1+e^{-2\pi k}}\\
N_{--}^{\text{out}}(k) ~ &= ~ \frac{N_{--}^{\text{in}}(k)}{1+e^{2\pi k}}+\frac{N_{++}^{\text{in}}(k)}{1+e^{-2\pi k}} \, ,
\end{align}
where $N^{\text{in}}_{++}$ and $N^{\text{in}}_{--}$ are the in-going number densities from either side of the potential. We see that indeed the scattered pulse emerges with thermal factors $1+e^{\pm 2\pi k}$. For most of the radiated spectrum to actually be thermal, we see that $N^{\text{in}}_{++}$ and $N^{\text{in}}_{--}$ must be constant over a large range of energies. This was observed to be the case in the context of 2d string theory, starting from a coherent pulse, seen as an excitation over an appropriate Fermi-sea vacuum \cite{Schoutens:1993hu,Karczmarek:2004bw,Friess:2004tq}. In our context, since we do not yet have a first principles construction of the appropriate second quantised theory, this in-state may be chosen. For instance, a simple pulse with a wide-rectangular shape would suffice. One may hope to create such a pulse microscopically, by going to the second quantised description and creating a coherent state. Alternatively, one may hope to realize a matrix quantum mechanics model that realizes a field theory in the limit of large number of particles. After all, we know that each oscillator in our model really corresponds to a partial wave and not a single particle in the four dimensional black hole picture.

\paragraph{Second Quantization v/s Matrix Quantum Mechanics}
Given the quantum mechanical model we have studied in this article, we may naively promote the wave-functions $\psi_{lm}$ into fields to obtain a second quantized Lagrangian:
\begin{equation}
\mathcal{L} ~ = ~ \sum_{l,m} \, \int_{-\infty}^{\infty} \, du^\pm \, \psi_{lm}^\dagger\left(u^\pm,t\right) \left[i \partial_t  + \dfrac{i}{2} \left(u^\pm \partial_{u^\pm} + \partial_{u^\pm} u^\pm \right) + \mu_l \right] \, \psi_{lm} (u^\pm,t) \, .
\end{equation}
With a change of variables to go to Rindler coordinates, 
\begin{align}
&\psi_{lm}^{(\text{in/out})}(\alpha^\pm ,\rho^\pm,t) ~ = ~ e^{\rho^\pm/2} \psi_{lm}(u^\pm = \alpha^\pm e^{\rho^\pm},t) \, ,
\end{align}
the Lagrangian becomes relativistic
\begin{align}
\mathcal{L} ~ &= ~  \sum_{l,m} \, \int_{-\infty}^{\infty} \, d \rho^\pm \,  \sum_{\alpha^\pm=1,2} \, \Psi_{lm}^{\dagger (\text{in/out})} \left(\alpha^\pm ,\rho^\pm,t\right) \left(i \partial_t  - i  \partial_{\rho^\pm} + \mu_l\right) \Psi_{lm}^{(\text{in/out})}\left(\alpha^\pm ,\rho,t\right) \, ,
\end{align}
where the label `in' (out) corresponds to the $+$ ($-$) sign. The form of the Lagrangian being first order in derivatives indicates that the Rindler fields are naturally fermionic. In this description we have a collection of different species of fermionic fields labelled by the $\{l,m\}$ indices. And the interaction between different harmonics would correspond to interacting fermions of the kind above. The conceptual trouble with this approach is that each ``particle'' to be promoted to a field is in reality a partial wave as can be seen from the four-dimensional picture. Therefore, second quantizing this model may not be straight-forward \cite{Hooft:2016cpw}. It appears to be more appealing to think of each partial wave as actually arising from an $N$-particle matrix quantum mechanics model which in the large-$N$ limit yields a second quantized description. Since $N$ counts the number of degrees of freedom, it is naturally related to $c$ via 
\begin{equation}
\dfrac{1}{N^2} ~ \sim ~ c ~ = ~ \dfrac{8 \pi G}{R^2} ~ \sim ~ \dfrac{l^2_P}{R^2} \, .
\end{equation}
Therefore, $N$ appears to count the truly microscopic Planckian degrees of freedom that the black hole is composed of. The collection of partial waves describing the Schwarzschild black hole would then be a collection of such $N$-particle matrix quantum mechanics models. Another possibility is to describe the total system in terms of a single matrix model but including higher representations/non-singlet states to describe the higher $l$ modes. This seems promising because if one fixes the ground state energy of the lowest $l=0$ (or $l=1$ after antipodal) oscillator, the higher $l$ oscillators have missing poles in their density of states compared to the $l=0$, much similar to what was found for the adjoint and higher representations in
\cite{Boulatov:1991xz}. 
Finally we note that we can combine the chemical potential with the oscillator Hamiltonian to get
\begin{equation}
\hat{H}_{\text{tot}} ~ = ~ \sum_{l,m} \, \left[\dfrac{1}{2} \left(\hat{p}_{l m}^2 - \hat{x}_{l m}^2\right) + \dfrac{R^2}{8 \pi G} \left(\hat{L}^2+1\right)\right] \, ,
\end{equation}
with $\hat{L}^2=\sum_i \hat{L}^2_i$ giving the magnitude of angular momentum of each harmonic. One can then perform a matrix regularisation of the spherical harmonics following \cite{deWit:1988wri,Taylor:2001vb} which replaces the spherical harmonics $Y_{l m}(\theta, \phi)$ with $N \times N$ matrices $\mathbb{Y}_{l m}$ where $l \leq N -1$. This naturally sets a cut-off on the spherical harmonics from the onset. To sharpen any microscopic statements about the S-matrix, one might first need to derive an MQM model that regulates Planckian effects. 

\section*{Acknowledgements}
We are indebted to Gerard 't Hooft for passing on to us, his infectious enthusiasm for black hole physics, for many extremely important comments on various drafts of this article and several encouraging discussions. We are grateful to Jan Manschot for very illuminating correspondence and discussions. We also take pleasure in thanking Umut G\"{u}rsoy, Javi Martinez-Magan, Phil Szepietowski and Stefan Vandoren for their insightful comments. \\
This work is supported by the Netherlands Organisation for Scientific Research (NWO) under the VIDI grant 680-47-518, the VICI grant 680-47-603, and the Delta-Institute for Theoretical Physics (D-ITP) that is funded by the Dutch Ministry of Education, Culture and Science (OCW).

\appendix

%
%
\bibliographystyle{JHEP}
\bibliography{BHSMatrix_IHO}

\end{document}